\documentstyle[12pt]{article}
\input epsf

\begin{document}
\title {
\begin{flushright}
{\normalsize IIT-HEP-95/2\\
hep-ex/9505002
}
\end{flushright}
\vskip 0.2in
\bf An Ultrahigh-Statistics Charm Experiment for the Year
$\sim2000$\thanks{Presented at the LISHEP95 Workshop, Rio de Janeiro, Brazil,
February 20--22, 1995.}}
\author{ Daniel M. Kaplan and Vassili Papavassiliou
     \\ {\sl Illinois Institute of Technology, Chicago IL 60616} 
	\\
         }  
\date{April 1995}
\maketitle

\begin{abstract}

After reviewing the motivation for high-statistics charm studies, we
describe a fixed-target experiment capable of reconstructing $>10^8$ charm
decays. High-rate silicon and scintillating-fiber tracking systems
allow operation at a 5\,MHz interaction rate,
using an 800\,GeV primary proton beam incident on a
thin target. The trigger includes transverse energy and/or lepton requirements
and also requires evidence of decay vertices. Simulations show that adequate
rejection can be obtained.

\end{abstract}

\section{Charm Physics at Fermilab: Brief History and Outlook}

\begin{figure}[htb]
\hspace{-0.25 in}\centerline{\epsfysize=3 in\epsffile{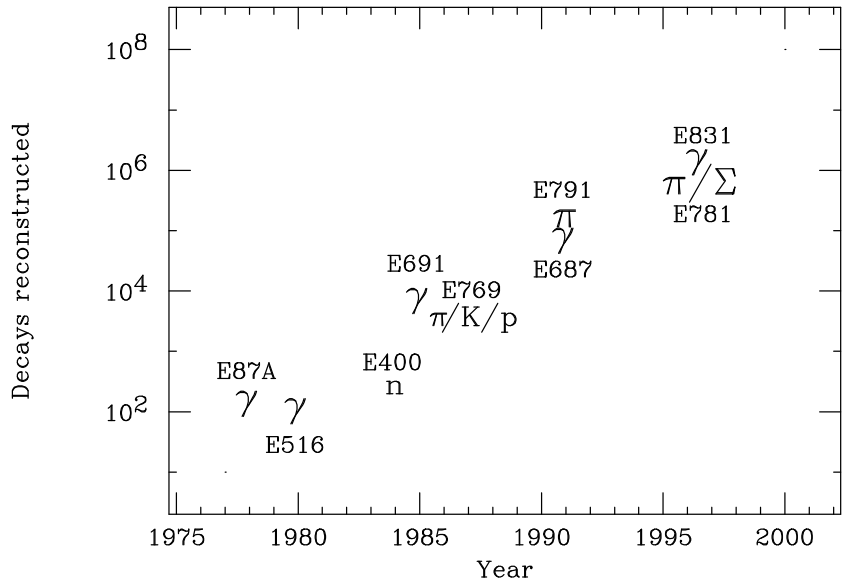}}
\caption{Yield of reconstructed charm vs.\ year of run
for completed and approved high-statistics
Fermilab fixed-target charm experiments; symbols indicate type of beam
employed.\label{history}}
\end{figure}

The program of fixed-target charm studies at Fermilab which began
in earnest about 1984
with E691 has been by any measure extraordinarily successful. The history of
Fermilab charm experiments is illustrated in
Fig.~\ref{history}~\cite{Christian-Appel}, which shows roughly exponential
growth in sensitivity since the late 1970s.
While the physics reach of such experiments depends both on the number of
signal events reconstructed and on the amount of background under the peaks,
the former figure can still serve as a starting point for discussion. This
number is expected to reach $\sim10^6$ events during the next few years with
the runs of Fermilab E781 and E831 (and the advent of CLEO III).

We are naturally led to the question whether further substantial  advances in
sensitivity can be foreseen in the years beyond. Our conclusion (detailed
below) is that an optimized fixed-target apparatus can reach $\sim10^8$
reconstructed decays using technology which is now or will soon be available.
At these levels of sensitivity charm studies will provide increasingly
incisive tests of the Standard Model and will also begin to probe
significantly beyond it~\cite{Sokoloff}.

\subsection{Searches for new physics}

Given the large number of free parameters in the Standard Model, we
must suspect that some new physics lies beyond it.
This new physics could have detectable effects in the charm sector.
For example,
Pakvasa~\cite{Pakvasa} has emphasized that rare and forbidden
processes such as $D^0 - \overline D {}^0 $
mixing, charm-changing neutral currents, and even
lepton-family-violating currents must occur at some level if we are ever to
understand the pattern of fermion masses and mixings.
While direct {\em CP} violation in singly-Cabibbo-suppressed (SCSD) charm
decays
is  expected~\cite{Burdman} in the Standard Model at the
$_\sim$\llap{$^<$}\,$10^{-3}$ level (possibly observable at the level of
sensitivity discussed here), {\em CP} violation in Cabibbo-favored
or doubly-Cabibbo-suppressed modes would
be a clear signature of new physics.

We can identify a growing trend towards precision studies of the strange
quark (e.g.\ E871 and the KTeV and KAMI programs at Fermilab and
E787 and E865 at the
Brookhaven AGS) and the beauty quark (CDF and D0 upgrades,
HERA-$B$, and the SLAC and KEK $B$
Factories). The major goal of these efforts  is the detailed study of
suppressed Standard-Model effects ({\em CP} violation and rare decays),
where small deviations may reveal aspects of the ultrahigh-energy
theory.
Various
proposed extensions of the Standard Model (e.g.\ multiple-Higgs-doublet
models with~\cite{Hall-Weinberg,tree}
or without~\cite{multiple-Higgs}
tree-level flavor-changing couplings,
left-right-symmetric models~\cite{left-right},
and models with family symmetry~\cite{family}
or supersymmetry~\cite{SUSY})  predict effects also in the charm sector which
should be detectable at the level of  sensitivity discussed here.
Charm experiments are uniquely sensitive to couplings of
these new currents to ``up-like" quarks, which may differ from those to
``down-like" quarks~\cite{up-down}. Furthermore, the Standard-Model
``backgrounds"
in processes such as particle-antiparticle mixing,
flavor-changing neutral currents, and {\em CP} violation are
far smaller in charm than in strangeness and beauty, resulting in a larger and
different discovery window.
Precision studies of charm are thus complementary to those of strangeness and
beauty.

$ D^0 - \overline D {}^0 $ mixing may be one of the more promising places to
look for low-energy manifestations of physics beyond the Standard Model.
Standard-Model contributions to $|\Delta m_D|$
are estimated~\cite{Burdman} to give $r_{\rm mix}<10^{-8}$.
Many nonstandard models predict a much larger effect.
An interesting example is the multiple-Higgs model of Hall and
Weinberg~\cite{Hall-Weinberg}, in which $|\Delta m_D|$ can be as large as
$10^{-4}$\,eV, approaching the current experimental limit.
In this model $K^0$ {\em CP} violation arises from the Higgs sector, and
{\em CP} violation in the beauty sector is expected to be small.

Charm-changing neutral currents
are forbidden at tree-level in the Standard
Model and thus are predicted~\cite{Schwartz}
to have branching ratios of order $10^{-11}$ to $10^{-8}$.
A variety of non-Standard models~\cite{FCNC} predict effects substantially
larger than this. Experimental sensitivities are now in the range $\sim10^{-5}$
and are expected to reach $\sim10^{-6}$ in E831.

In summary (based on discussions and presentations at the CHARM2000
Workshop~\cite{CHARM2000}),
we anticipate the following order-of-magnitude estimates for sensitivity to new
physics in a ``$10^8$-charm" experiment:
\begin{itemize}
\item $D^0$ mixing: $r_{\rm mix}\sim10^{-5}$

\item charm-changing neutral currents and
lepton-family-violating currents: $B\sim10^{-7}$

\item charm {\em CP} violation: $A_{CP}\sim10^{-3}$ in SCSD modes
\end{itemize}

\subsection{Tests of the Standard Model}

Incisive tests of the CKM Model are envisioned from precision beauty
studies at CLEO, the Tevatron Collider, HERA-$B$, and the $B$ factories.
Here too, high-sensitivity charm-quark studies play an important
role complementary to
that of $b$ experiments:

\begin{enumerate}
\item Several authors~\cite{Burdmanetal,Rosner} have noted that suppressed
Standard-Model processes (including such nonperturbative effects as
final-state interactions) can complicate attempts to extract CKM parameters
from beauty measurements. Charm decay provides an ideal laboratory in which to
study these effects, which are both larger (due to the smaller charm-quark
mass) and more cleanly studied (due to the small size of Standard-Model {\em
CP}-violating and penguin-mediated effects) in charm than in beauty.
In addition, the interplay of perturbative and nonperturbative Standard-Model
physics at work here is of  interest in its own right~\cite{Luke}.

\item The extraction of CKM parameters from beauty measurements in many cases
requires knowledge of  form factors and decay constants. These have been
predicted using various models, including lattice gauge calculations and
calculations employing the heavy-quark symmetry of  QCD. The
predictions of these models
need to be tested in the charm sector to verify their applicability in the
beauty sector.

%
\item Charm is the lightest quark for which perturbative QCD calculations may
be meaningful, thus its production in large numbers in hadronic
interactions can provide detailed tests of QCD. Understanding the production
mechanism in terms of perturbative-QCD processes could lead to a determination
of the gluon distribution of the proton at $x\,_\sim$\llap{$^>$}\,0.025,
complementary to the small-$x$ measurements possible at HERA and the
hadron-collider experiments. The interplay of perturbative and
non-perturbative effects here presents the opportunity to study longer-range
aspects of QCD, such as the potential existence (and magnitude) of an
intrinsic charm component in the nucleon, and final-state interactions
in propagation of charm quarks through nuclear matter.
%
\end{enumerate}

\section{A Next-Generation Charm Spectrometer}

We next discuss a hypothetical experiment which potentially can achieve the
$10^8$-reconstructed-charm sensitivity mentioned above. As we will see, the
most demanding requirement is on the trigger. In particular, an on-line
secondary-vertex trigger is needed if adequate trigger rejection is to be
achieved without sacrificing sensitivity in  hadronic decay modes.

\subsection {Beam and target}

To achieve  the sensitivity discussed here in a fixed-target run of
$\approx10^5$ beam spills requires a primary
proton beam (see Table 1~\cite{architecture}). Given
$\sigma\,(pN\to D\,X) + \sigma\,(pN\to {\overline D {}} \,X) \approx
40\,\mu$b/nucleon at 800\,GeV~\cite{world-average} and  $\sigma\,(pN\to
D^0\,X)\propto A^{1.0}$~\cite{Leitch}, and assuming that the cross section to
produce $D_s$ and charmed baryons is $\approx15$\% that of $D$ mesons, the
charmed-particle production rate is
$7\times10^{-3}$/interaction if a high-$A$ target (e.g.\ Au) is used.

\begin{table}[htb]
\caption{Charm yields$^*$ for various possible beams using an open-geometry
general-purpose spectrometer. \label{tab:beam}}
\begin{center}
\begin{tabular}{|l|c|c|c|c|}
\hline
Beam & $\sigma_{c \bar{c}}$ & Max.\ beam& Charm events& Reconstructable\\
 & (cm$^2$) & /spill &/spill & charm/spill${}^{\footnotesize\dag}$\\
\hline
\hline
p (800 GeV) & $2\times10^{-29}$ & $10^{13}$ &
$7\times10^4\,{}^{\footnotesize\ddag}$&
$\approx1.5\times10^3\,{}^{\footnotesize\ddag}$\\
$\pi^-$ (500 GeV) & $2\times10^{-29}$ & $10^{8}\,{}^{\footnotesize\S}$ &
$3\times10^3$ & $\approx60$\\
$\gamma$ ($\approx$200 GeV) & $1\times10^{-30}$ &
$10^{8}\,{}^{\footnotesize\S}$ & $2\times10^2$ & $\approx4$\\
\hline
\end{tabular}
\end{center}
{\footnotesize
\begin{list}{}{\setlength{\itemsep}{0pt}\setlength{\parsep}{0pt}}
\item[$^*$]neglecting $A$-dependence enhancement.
\item[${}^{\dag}$]based on $\Sigma$\,(branching ratio $\times$
efficiency)$\,\approx10^{-2}$ for  reconstructable modes.
\item[${}^{\ddag}$]assuming a maximum of 0.1 interaction/bucket.
\item[${}^{\S}$]limited by proton economics.
\end{list}
}
\end{table}

A target which is short compared to typical charm decay lengths is crucial for
optimizing background suppression, both off-line and at trigger level. While
multiple thin targets could be employed (as in E791 and E831),
a single  target facilitates fast vertex triggering. A 1\,mm W, Pt, or Au
target is one possibility, representing $\approx$1\% of an interaction length
and on average 14\% of a radiation length for outgoing secondaries. A low-$Z$
material  such as $^{13}$C-diamond may be favored to minimize scattering of
low-momentum pions from $D^*$ decay~\cite{architecture};
then a 2\,mm  target is suitable,
representing $\approx$1\% of an interaction length and $\approx$1\% of a
radiation length and producing charm at the rate
$3\times10^{-3}$/interaction. Given the typical Lorentz boost $\gamma\approx
20$,  a 1--2\,mm target is short enough that a substantial fraction even of
charmed baryons will decay outside it.

We consider a benchmark 5\,MHz interaction rate, which then requires
0.5--1\,GHz of primary proton beam, an intensity easily attainable.
As shown below, this
yields  $_\sim$\llap{$^>$}\,$10^8$ reconstructed charm per
few\,$\times\,10^6$\,s of
beam ($\approx10^5$ spills $\times~20\,$s/spill).

\subsection {Rate capability}

\begin{figure}[htb]
\centerline{\epsfysize = 2.1 in \epsffile {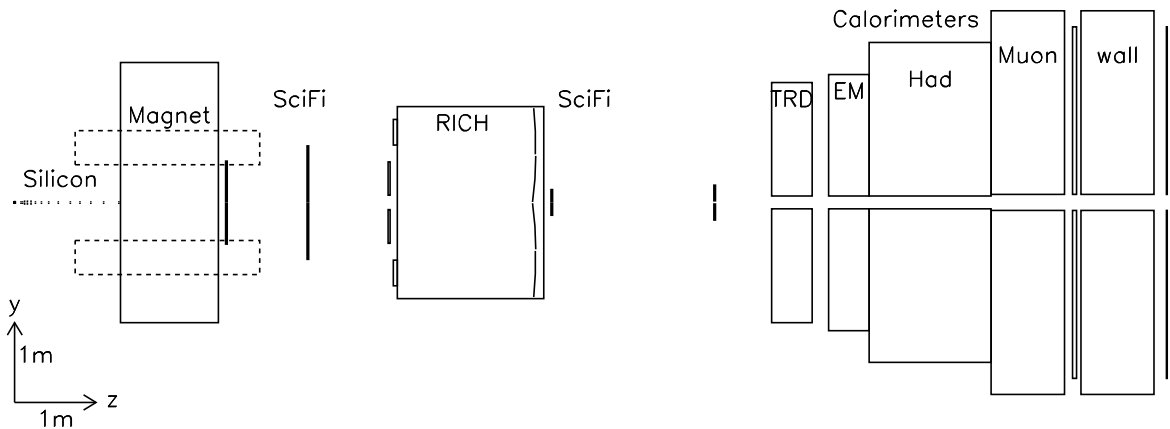}}
\caption [Spectrometer layout (bend view).]%
{Spectrometer layout (bend view).}
\label{app865}
\end{figure}

\begin{figure}[htb]
\centerline{\epsfysize = 1.8 in \epsffile {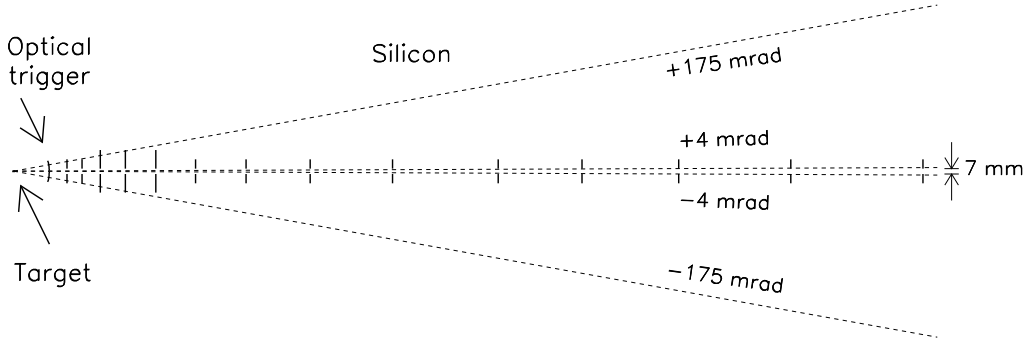}}
\caption [Detail of vertex region (showing optional optical impact-parameter
trigger).]%
{Detail of vertex region (showing optional optical impact-parameter
trigger).}
\label{detail865}
\end{figure}

A significant design challenge is posed by radiation damage to the silicon
detectors. To configure detectors which can survive at the desired
sensitivity, we choose suitable maximum and (in one view) minimum angles for
the instrumented aperture, arranging the detectors along the beam axis with a
small gap through which pass the uninteracted beam and secondaries below the
minimum angle (Figs.~\ref{app865},~\ref{detail865}).\footnote{An
alternative approach with no gap may also be workable if the beam is spread
over sufficient area to satisfy rate and radiation-damage limits, however the
approach described here
probably allows smaller silicon detectors and is ``cleaner" in that the beam
passes through a minimum of material.} Thus the rate is spread approximately
equally over several detector planes, with
large-angle secondaries measured close to the target and small-angle
secondaries farther downstream. Along the beam axis the spacing of detectors
increases geometrically, making the lever arm for vertex
reconstruction independent of production angle. Since small-angle secondaries
tend to have high momentum, the multiple-scattering contribution to vertex
resolution is also approximately independent of production angle. We have
chosen an instrumented angular range $|\theta_x| \leq 200\,$mr, $4\leq
\theta_y \leq 175\,$mr, corresponding to the center-of-mass rapidity range
$|y|\,_\sim$\llap{$^<$}$\,1.9$ and containing over 90\% of produced
secondaries.

To maximize the rate capability of the spectrometer, tracking is performed
entirely with silicon and scintillating-fiber planes.
The rate per unit area (and hence the radiation fluence) in a detector element
can easily be estimated based on the uniform-pseudorapidity approximation.
Fig.~\ref{rate} shows the rate calculation for
an annular area $dA$ located a transverse distance $r$ from
the beam and of thickness $dr$.
Since the operational limit of present-day silicon detectors is
$10^{14}\,$particles/cm$^2$, the charged multiplicity per unit
pseudorapidity in 800\,GeV proton-nucleus collisions
is $n\approx4$ for
high-$A$ targets~\cite{HERA-B} (less for C),
and calling the total number of interactions in the run
$n_{\rm int}$,
we can derive the ``minimum survivable"
inner detector radius
$$r_{\rm min}= {\left ( {n\over2\pi}{n_{\rm int}\over{10^{14}}}\right )}
^{1\over2}\,.$$
Since a typical run will yield fewer than
$(5\times10^6\,{\rm interactions/s})\times(4\times10^6\,{\rm s})
=2\times10^{13}$
interactions, we obtain conservatively
$r_{\rm min}=
3.5$\,mm, which we set as the half-gap between the two detector
arms. This ensures that the detectors will survive for the entire run (or
at most will
need to be replaced once\footnote{In E789 we operated silicon
detectors
at fluence up to $\approx5\times10^{13}\,$cm$^{-2}$ with negligible
efficiency loss.}).
To cover the desired angular range, we configure 14 double-sided
silicon strip detectors\footnote{We assume silicon strip detectors for
definiteness, but silicon pixel detectors would be better if available
with sufficient readout speed and radiation hardness; because of their
radiation hardness, diamond
detectors~\cite{Tesarek}, if available, should also be considered.}
above and 14 below the beam as
shown in Fig.~\ref{detail865}, such that at all angles of interest there are
at least six measurements per track (and more at small angles where the
occupancy is highest).

\begin{figure}[htb]
\centerline{\epsfysize = 1.5 in \epsffile {rate_fig.epsf}}
\caption [Calculation of rate per unit area in an annulus.]%
{Calculation of rate per unit area in an annulus.}
\label{rate}
\end{figure}

A large-scale scintillating-fiber tracker has recently been successfully
tested~\cite{Ruchti}
for the D0 upgrade, and we base our fiber-tracker
design on that demonstrated technology.
Green-scintillating 3HF/PTP
fibers are deployed in staggered doublets in three views.
They are read out using cryogenic
solid-state ``visible-light photon counters" (VLPCs) \cite{Atac3}.
Since the
fibers are more radiation-hard than silicon
detectors, and (due to occupancy; see below)
the beam gap between fiber planes
is larger than that in the silicon,  radiation damage of the fibers is not
anticipated to be a problem.

We assume 1-bucket ($<$19\,ns) recovery times for all detectors, so that
there is no pile-up due to out-of-time interactions. Designs capable of this
performance have been presented~\cite{HERA-B,Bari} for all detectors
except the TRD.\footnote{It
may be that a TRD for electron identification is not cost-effective and
a hadron-blind detector~\cite{HBD} or preshower detector should be used
instead.}

Detector-element occupancies also follow from the derivation of
Fig.~\ref{rate}.
For an element of height $dy$ located a transverse
distance $y$ from the beam and covering $-x_{\rm max}<x<x_{\rm max}$,
the occupancy per event (neglecting magnetic bending) is
$${n\over\pi}{dy\over y}\arctan{x_{\rm max}\over y}\,.$$ For 800\,$\mu$m fiber
diameter, this implies $\approx$16\% occupancy at $y=1\,$cm,
$\approx$8\% at 2\,cm, and $\approx$4\% at
4\,cm. A full trackfinding
simulation will be required to assess the maximum acceptable occupancy,
but this suggests $\approx$1\,cm as the minimum acceptable half-gap in the
scintillating-fiber planes. The fibers near the gap could be split at
$x=0$ and read out at both ends, halving their occupancies. Since shorter
fibers have less attenuation, a smaller diameter could be used near the
gap, reducing occupancy still further.

\subsection{Spectrometer performance}

We have carried out a simple
Monte Carlo simulation of the spectrometer sketched above.
Assuming a 1.2-m-long analyzing magnet with pole pieces
tapered to give 0.5\,GeV $p_t$ kick, we obtain ($56\pm1$)\%
geometrical acceptance for $D^0\to K^-\pi^+$ decays and ($44\pm1$)\%
for $D^{*+}\to D^0\pi^+\to K^-\pi^+\pi^+$, comparable to those of
existing open-geometry spectrometers despite the beam gap.
With silicon detectors of 25\,$\mu$m pitch read out digitally (i.e.\ no
pulse-height information) and 800\,$\mu$m scintillating-fiber pitch, and
assuming $\pm10^\circ$ stereo, Gaussian fits to the reconstructed distributions
give rms resolutions of 6\,MeV in mass  (a
factor $\approx2$ better than that of existing spectrometers) and 11\,$\mu$m
(bend-view) and 21\,$\mu$m (nonbend-view) in impact parameter, giving 40\,fs
decay proper-time resolution, comparable to that of existing spectrometers.
Since the mass resolution is dominated by scattering,
minimization of material is crucial,
for example use of helium bags and avoidance of threshold Cherenkov counters
employing heavy gas mixtures.
(The performance parameters just given
are a snapshot of work in progress and probably can be
improved with further optimization.)

\subsection{Trigger}

While the most successful previous charm hadroproduction experiments (E769 and
E791) used very loose triggers and recorded most inelastic interactions, this
approach is unlikely to extrapolate successfully by three orders of magnitude!
(Consider that E791 recorded $2\times10^{10}$ events -- tens of terabytes of
data -- on 20,000 8\,mm tapes.) Thus our sensitivity goal requires a highly
selective trigger.
However, we wish to trigger on charm-event characteristics
which bias the physics as little as possible.
Lepton triggers, used
successfully by E653, while capable of great selectivity ($\sim10^3$
rejection for minimum-bias events), have only $\sim10$\% charm efficiency.
The $E_t$ triggers used by E769 and E791, while highly efficient for charm,
have poor selectivity ($_\sim$\llap{$^<$}$\,10$ minimum-bias rejection).
We therefore assume a first-level trigger requiring
calorimetric $E_t$ OR'ed with high-$p_t$-lepton and lepton-pair triggers.
At second level, secondary-vertex requirements can be imposed
on the $E_t$-triggered events to achieve a rate
($\sim100\,$kHz) which is practical to record.

\begin{figure}[htb]
\centerline{\epsfysize = 3.5 in \epsffile {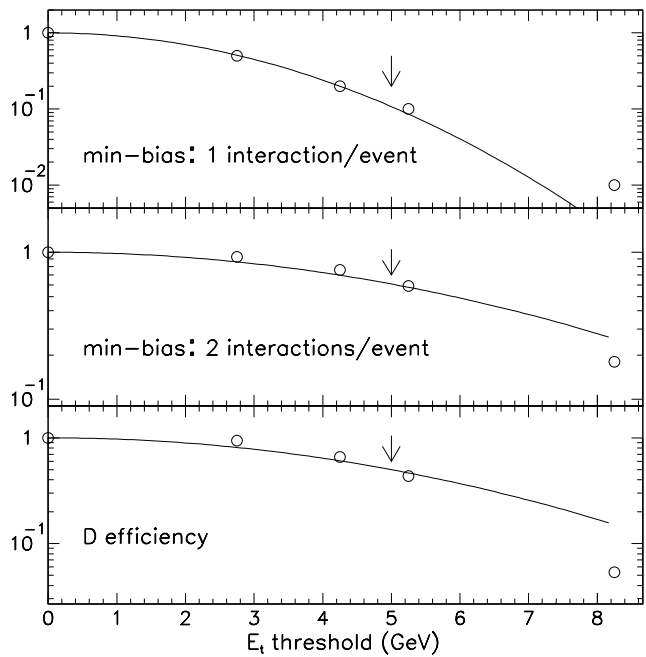}}
\caption [Minimum-bias and $D$ efficiencies vs.\ $E_t$ threshold.]%
{Minimum-bias and $D$ efficiencies vs.\ $E_t$ threshold;
points are from PYTHIA simulation of Ref.~\cite{Karchin},
curves are fits of the form $\exp{(-aE_t^2)}$, and arrows indicate
the 5\,GeV ``PYTHIA $E_t$" threshold discussed in text.}
\label{Et-optrig}
\end{figure}

Analyses of
the efficacy of an $E_t$ trigger carried out using E791
data~\cite{P829,Christian} and the PYTHIA Monte Carlo~\cite{Karchin}
agree on minimum-bias rejection vs. charm
efficiency (though due to nuclear effects not simulated in
PYTHIA, they differ as to the $E_t$ threshold corresponding to a given
rejection). Fig.~\ref{Et-optrig}
shows the efficiencies for charm and minimum-bias
events as a function of the PYTHIA $E_t$ threshold. A considerable
degradation results if there is significant probability for two interactions
to pile up in the calorimeter. Given the 53\,MHz rf structure of the Tevatron
beam and the typical $\approx50$\% effective spill duty factor, at the
benchmark 5\,MHz mean interaction rate there is a $\approx20$\% probability
for a second simultaneous interaction. Thus at a 5\,GeV PYTHIA $E_t$
threshold (corresponding to a $\approx$10\,GeV actual
threshold),
the minimum-bias rejection factor is 5, i.e.\ pile-up degrades the
rejection by a factor $\approx2$, even for a calorimeter with one-bucket
resolution. The charm efficiency at this threshold
is about 50\%, for a charm enrichment of
$\approx2.5$. (These are rough estimates based on a relatively crude
calorimeter~\cite{Karchin}, and an optimized calorimeter may provide better
rejection.) Such an $E_t$ trigger yields a 1\,MHz input rate to the next level.

While it may be technically feasible by the Year 2000 to record events at a
1\,MHz rate, an additional factor $\approx$10 in trigger rejection is
desirable, and can be achieved by requiring evidence of secondary vertices.
Existing custom trackfinding trigger processors~\cite{Processors},
while perhaps capable of this rejection, typically fall short by $\approx$ one
order of magnitude in speed. At $\sim 1\,$MIPS-s/event,
an on-line farm of commercial processors would need a capacity of
$\sim10^5~-~10^6\,$MIPS, which may be prohibitive even in the Year 2000.
It is likely that by then a sufficiently fast
custom trackfinding processor
can be developed. This would require fast buffering ($\sim100\,$ns) and
readout ($\sim1\,\mu$s) of event information in order not to impose excessive
deadtime. Trackfinding secondary-vertex triggers benefit from the use of
focused beam and a single
thin target, which allow simplification of the algorithm
since the primary vertex location is known {\it a priori}. Since low-$p_t$
tracks have poor vertex resolution~\cite{Selove}, a
trigger which discriminates $p_t$ is more effective than one which is purely
topological;
such discrimination may be simply accomplished by placing the
vertex detectors in a weak magnetic field and looking for straight
tracks. Christian~\cite{Christian} has suggested a simple trigger-processor
algorithm based on this idea.

\begin{figure}[htb]
\vspace{-.5 in}
\hbox {
\hspace{.75 in}
{\epsfysize = 5 in \epsffile {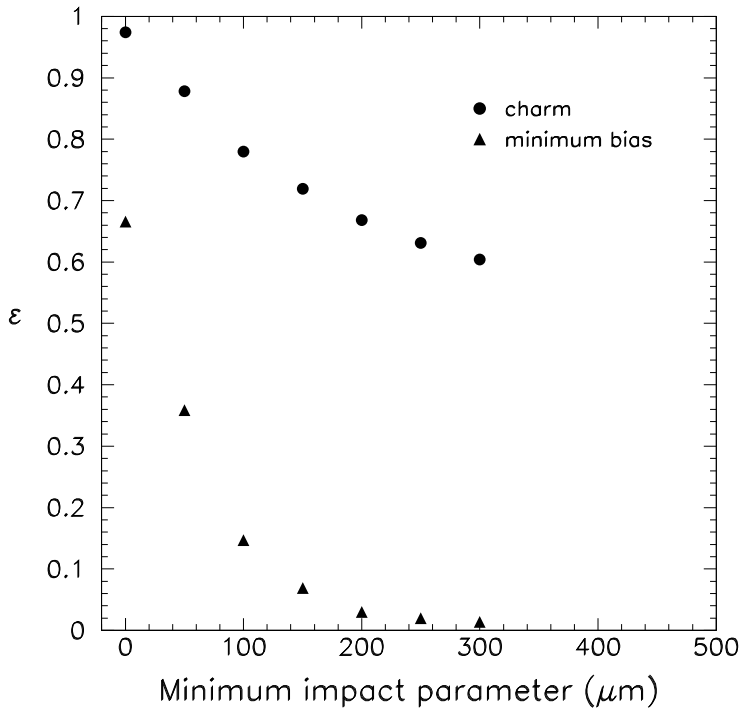}}
\vspace{-.75 in}
\hfill}
\caption {Efficiency of the vertex-trigger algorithm as a function of
the minimum impact parameter for charm events (dots) and minimum-bias
events (triangles), using a PYTHIA-based Monte Carlo simulation.}
\label{vxtrig}
\end{figure}

%
As a first step in evaluating the efficacy of such a trigger,
a Monte Carlo simulation was performed for the silicon-detector
configuration shown in
Fig.~\ref{detail865}, using PYTHIA for the event generation. The target
length was taken to be 1\,mm, while its transverse dimensions were ignored,
approximating a very narrow target (at least in one view).
25\,$\mu$m strip pitch was assumed.
An algorithm was applied that searched for hits in two seed planes in
the $y$-$z$ view. When a pair of hits was found with more than
some minimum impact parameter when extrapolated to the target, additional $y$
hits,
consistent with the line defined by the two seed planes, were sought.
The search was repeated using different pairs of seed planes, and an event
was accepted if at least one track was found. Fig.~\ref{vxtrig} shows the
efficiency of the algorithm, both for charm and for minimum-bias events,
as a function of the minimum impact parameter. It can be seen that for an
impact-parameter cut of 200\,$\mu$m, a rejection factor of more than 20 can be
obtained for the background, while retaining a 67\% efficiency for charm
events.
Somewhat surprisingly,
the addition of a weak magnetic field
(up to 3 kG) in the target region was found not to be
helpful: because the algorithm looks for straight tracks,
a field strong enough to improve the rejection degrades the efficiency
significantly.

\begin{figure}[htb]
\vspace{-.5 in}
\hbox {
\hspace{.75 in}
{\epsfysize = 5 in \epsffile {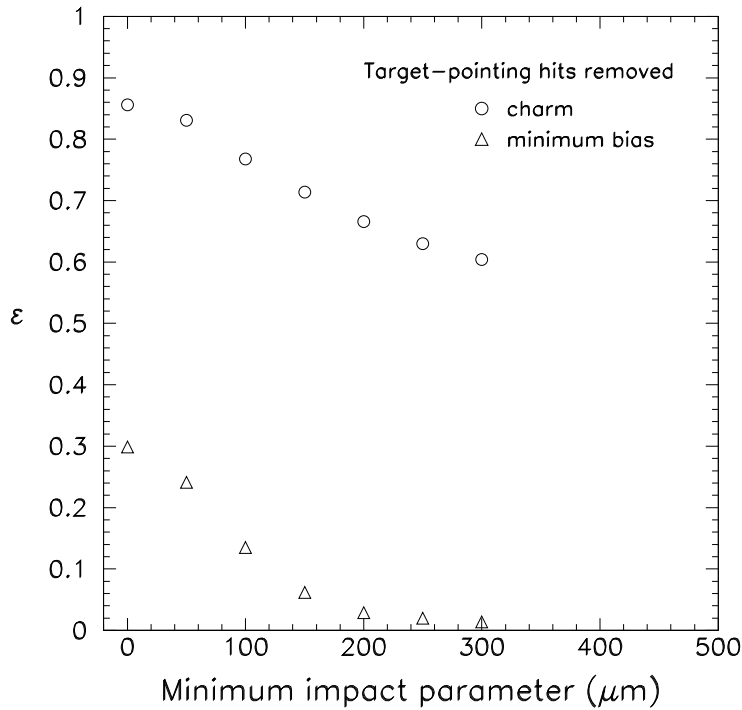}}
\vspace{-.75 in}
\hfill}
\caption {Efficiency of the vertex-trigger algorithm as a function of
the minimum impact parameter for charm events (dots) and minimum-bias
events (triangles), using a preliminary hit-elimination pass to speed up
processing.}
\label{vxtrig2}
\end{figure}

A key feature of Christian's algorithm is the elimination of hits which lie on
straight lines pointing to the target prior to searching for tracks of finite
impact parameter. This can reduce substantially the processing time, since as
the number of hits per detector plane ($n$) increases, the
time to eliminate hits is linear in $n$, while the time to find tracks of
finite impact parameter goes as $n^2$ (due to the loops over hits in two seed
planes)~\cite{Christian}. The results using this preliminary hit-elimination
pass are shown in Fig.~\ref{vxtrig2}. For the 200\,$\mu$m impact-parameter cut
discussed above, the rejection and efficiency are hardly affected. This
algorithm thus looks quite promising.

%

\begin{figure}[htb]
\centerline{\epsfysize = 4 in \epsffile {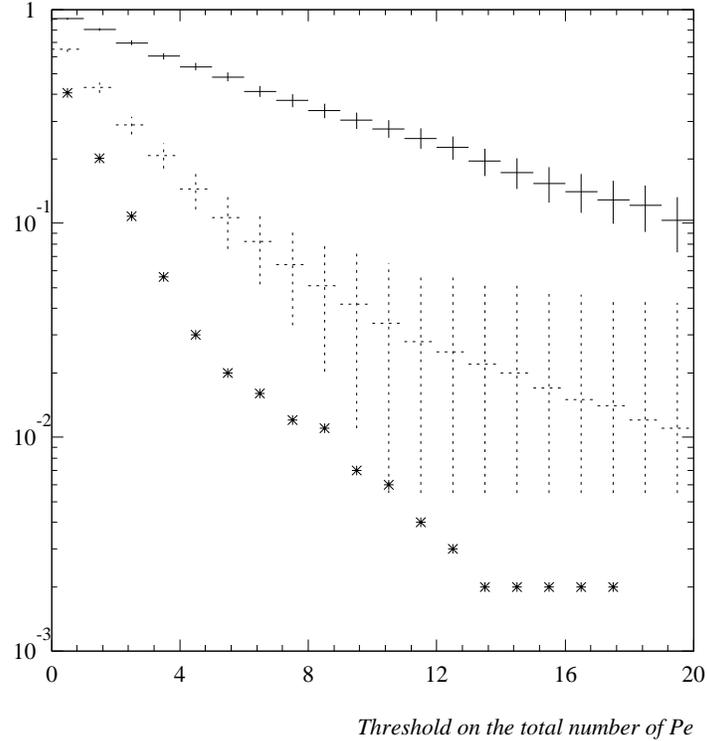}}
\caption [optical trigger performance]%
{Estimated optical-trigger efficiency for minimum-bias (solid crosses),
PYTHIA charm (dashed crosses), and
$B^0\to\pi\pi$ events (stars) vs.\ threshold in photoelectrons.}
\label{fig:optrig}
\end{figure}

As alternatives to iterative trackfinding at a 1\,MHz event rate,
three other approaches also appear worth pursuing.  The first is a
secondary-vertex trigger implemented using fast parallel logic, e.g.\ PALs,
neural networks, or pre-downloaded fast RAMs, to look
quickly for patterns in the silicon detectors
corresponding to tracks originating downstream of the target. The others are
fast secondary-vertex trigger devices originally proposed for beauty: the
optical impact-parameter~\cite{optrig} and Cherenkov
multiplicity-jump~\cite{mul-jump} triggers;
while results from prototype tests so far suggest
undesirably low charm efficiency, these might with further development provide
sufficient resolution to trigger efficiently on charm. For example,
Fig.~\ref{fig:optrig} shows the efficiency for minimum-bias, charm, and beauty
events projected for a version of the optical trigger~\cite{P865},
indicating 40\% charm
efficiency for a factor 5 minimum-bias rejection. The resulting
$\approx$200\,kHz event rate can be processed on-line or recorded using
existing technology.

\section{Yield}

The charm yield is straightforwardly estimated. Assuming a Au target and a
typical fixed-target run of $3\times10^6$ live beam seconds, $10^{11}$ charmed
particles are produced. The reconstructed-event yields in
representative modes are estimated
in Table~\ref{Yields} assuming (for the sake of illustration)
that the optical trigger is used for all-hadronic
modes (but not for leptonic modes, for which the first-level
trigger rate should be sufficiently low to be
recorded directly) and performs as estimated
above. Although due to off-line selection cuts not yet simulated,
realistic yields could be a factor $\approx2-3$ below those indicated,
the total reconstructed sample is
well in excess of $10^8$ events. Given the factor $\approx2$
mass-resolution improvement
compared to E791, one can infer a factor $\sim 50$ improvement in statistical
significance in a typical decay mode. Since the charm cross section at 120\,GeV
proton-beam energy may be several \% of that at
800\,GeV, and the geometrical acceptance remains $\approx50$\%,
interesting charm sensitivity may also  be available using Main Injector beam
during Tevatron Collider running; at the least, there will be opportunity to
debug and test the spectrometer thoroughly so that full-energy beam may be
used with optimal efficiency.

\begin{table}\centering
\caption
{ Estimated yields of reconstructed events (antiparticles included)}
\label{Yields}
\vspace{5mm}
\begin{tabular}{|l|c|c|c|c|c|}
\hline
mode & charm frac. &
BR & acceptance & efficiency & yield \\
\hline
\hline
$D^0\to K\pi$ & 0.6 & 0.0365 & 0.56 & 0.1 & $1.2\times10^8$ \\
$D^+\to K^*\mu\nu$ & 0.3 & 0.027 & 0.4 & 0.25 & $8\times10^7$\\
\qquad$\to K\pi\mu\nu$ &&&&&\\
all & 1 & $\approx0.1$ & $\approx 0.4$ & $\approx 0.1$ &
$\approx4\times10^8$\\
\hline
\end{tabular}
\end{table}

\section{Summary}

A fixed-target hadroproduction experiment capable of reconstructing in
excess of $10^8$ charm events is feasible using detector, trigger, and data
acquisition technologies which exist or are under development.
A typical factor $\sim50$ in statistical
significance of signals may be expected compared to E687 and E791.
We expect the spectrometer sketched here to cost substantially less than
HERA-$B$ (whose cost was estimated at 33M DM in 1994).
We anticipate an exciting future for charm physics at the
turn of the century.

\section*{Acknowledgements}

We have benefited from discussions with C. N. Brown, D. C. Christian, S. Kwan,
M. D. Sokoloff, and S. Pakvasa, and we thank them for reading the manuscript.
We thank the organizers for a most stimulating workshop.


\begin{thebibliography}{999}

\bibitem{Christian-Appel}
After a figure prepared by  by D. C. Christian for Fermilab Proposal 829
and included in J. A. Appel, ``Experimental Issues in High-Sensitivity Charm
Physics,"  in {\bf The Future of High-Sensitivity Charm
Experiments}, {\sl Proceedings of the CHARM2000 Workshop},
Fermilab, June 7--9, 1994,
D. M. Kaplan and S. Kwan, {\it eds.}, FERMILAB-Conf-94/190, p. 1.

\bibitem{Sokoloff}
M. D. Sokoloff and D. M. Kaplan, ``Physics of an Ultrahigh-Statistics Charm
Experiment," IIT-HEP-95/1, to appear in {\sl Proc.\ of the HQ94 Workshop},
Charlottesville, VA, Oct.\ 7--10, 1994.

\bibitem{Pakvasa}
S. Pakvasa, ``Charm as Probe of New Physics,"
in {\bf The Future of High-Sensitivity Charm
Experiments}, {\it op cit.}, p.\ 85.

\bibitem{Burdman}
G. Burdman, ``Charm Mixing and CP Violation in the Standard Model,"
in {\bf The Future of High-Sensitivity Charm
Experiments}, {\it op cit.}, p.\ 75.

\bibitem{Hall-Weinberg}
Lawrence Hall and Steven Weinberg, Phys.\ Rev.\ D {\bf 48}, R979 (1993).

\bibitem{tree}
S. Pakvasa and H. Sugawara, Phys.\ Lett.\ {\bf 73B}, 61 (1978);
M. Shin, M.
Bander, and D. Silverman, in {\sl Proc.\ Tau-Charm
Factory Workshop},  Stanford, CA, May 23--27, 1989, SLAC-Report-343, p.\ 686.

\bibitem{multiple-Higgs}
L. F. Abbott {\it et al.}, Phys.\ Rev.\ D {\bf 21}, 179 (1980);
V. Barger {\it et al.}, Phys.\ Rev.\ D {\bf 41}, 3421 (1990).

\bibitem{left-right}
A. Le Yaouanc {\it et al.}, Phys.\ Lett.\ B {\bf 292}, 353 (1992);
M. Gronau and S. Wakaizumi, Phys.\ Rev.\ Lett.\ {\bf 68}, 1814 (1992).

\bibitem{family}
G. Volkov {\it et al.}, Yad.\ Fiz.\ {\bf 34}, 435 (1981).

\bibitem{SUSY}
A. Datta, Phys.\ Lett.\ {\bf 154B}, 287 (1985);
I. Bigi {\it et al.}, Z.\ Phys.\ {\bf C48}, 633 (1990);
Y. Nir and N. Seiberg, Phys.\ Lett.\ B {\bf 309}, 337 (1993).

\bibitem{up-down} A. Hadeed and B. Holdom, Phys.\ Lett.\ {\bf 159B}, 379
(1985); W. Buchmuller and  D. Wyler, Phys.\ Lett.\ {\bf 177B}, 377 (1986); W.
Buchmuller, in {\sl Proc.\  Int.\ Symp.\ on Production and Decay of Heavy
Hadrons}, Heidelberg, Germany, May 20--23, 1986, p.\ 299.

\bibitem{Schwartz}
K. S. Babu {\it et al.},
Phys.\ Lett.\ B {\bf 205}, 540 (1988);
A. J. Schwartz, Mod.\ Phys.\ Lett.\ {\bf A8}, 967 (1993).

\bibitem{FCNC}
W. Buchmuller and D. Wyler, Phys.\ Lett.\ {\bf 177B}, 377 (1986);
A. S. Joshipura, Phys.\ Rev.\ D {\bf 39}, 878 (1989);
Miriam Leurer, Phys.\ Rev.\ Lett.\ {\bf 71}, 1324 (1993).

\bibitem{CHARM2000}
D. M. Kaplan and S. Kwan, {\it eds.},
{\bf The Future of High-Sensitivity Charm
Experiments}, {\sl Proceedings of the CHARM2000 Workshop},
Fermilab, June 7--9, 1994,
FERMILAB-Conf-94/190; see especially
R. Morrison, ``CHARM2000 Workshop Summary," {\it ibid.}, p.\ 313.

\bibitem{Burdmanetal}
G. Burdman, E. Golowich, J. L. Hewett, and S. Pakvasa,
``Radiative Weak Decays of Charm Mesons," FERMILAB-Pub-94/412-T.

\bibitem{Rosner}
J. Rosner, ``What Charm Can Tell Us About Beauty,"
in {\bf The Future of High-Sensitivity Charm
Experiments}, {\it op cit.}, p. 297.

\bibitem{Luke}
See for example
M. Luke, ``Inclusive Charm Decays from QCD," in {\bf The Future of
High-Sensitivity Charm Experiments}, {\it op cit.}, p. 67.


\bibitem{architecture}
Adapted from Table 1 of C. N. Brown, D. M. Kaplan, and D. J.
Summers, ``Report of
the Working Group on Beams and Architectures," in {\bf The
Future of High-Sensitivity Charm Experiments}, {\it op cit.}, p. 459.

\bibitem{world-average} We average together measurements of
charged- and neutral-$D$ production from
R. Ammar {\it et al.}, Phys.\ Rev.\ Lett.\ {\bf 61},
2185 (1988); K. Kodama {\it et al.}, Phys.\ Lett.\ B {\bf 263}, 573 (1991);
and Ref.~\cite{Leitch}.

\bibitem{Leitch} M. J. Leitch {\it et al.},
Phys.\ Rev.\ Lett.\ {\bf 72}, 2542 (1994).

\bibitem {HERA-B}
T. Lohse {\it et al.}, ``HERA-$B$:
An Experiment to Study CP Violation in the $B$
System Using an Internal Target at the HERA Proton Ring," Proposal to DESY,
DESY-PRC 94/02, May 1994.

\bibitem{Ruchti}
R. Ruchti, ``Fiber Tracking," in {\bf The
Future of High-Sensitivity Charm Experiments}, {\it op cit.}, p.\ 173.

\bibitem{Atac3} M. D.
Petroff and M. Atac, IEEE Trans. Nucl. Sci. {\bf 36} (1989) 163;
M. Atac {\it et al.}, Nucl. Instr. \& Meth. {\bf A314} (1992) 56;
M. Atac {\it et al.}, Nucl. Instr. \& Meth. {\bf A320} (1992) 155.

\bibitem{Tesarek}
R. Tesarek, ``Diamond Detectors," in {\bf The
Future of High-Sensitivity Charm Experiments}, {\it op cit.}, p.\ 163.

\bibitem{Bari}
See e.g.\ D. M. Kaplan {\it et al.},
Nucl.\ Instr.\ \& Meth.\ {\bf A343}, 316 (1994);
D. F. Anderson, S. Kwan, and V. Peskov, {\it ibid.}, p. 109;
N. S. Lockyer {\it et al.}, {\it ibid.} {\bf A332}, 142 (1993).

\bibitem{HBD}
Y. Giomataris and G. Charpak,
Nucl. Instr. \& Meth. {\bf A310}, 589 (1991);
M. Chen {\it et al.},
``Results of a First Beam Test of Hadron Blind Trackers",
CERN-PPE-93-191, Aug. 1993.

\bibitem{P829}
J. C. Anjos {\it et al.}, ``Continued Study of Heavy Flavors at TPL,"
Fermilab Proposal 829 (1994).

\bibitem{Christian}
D. C. Christian, ``Triggers for a High-Sensitivity Charm Experiment,"
in {\bf The
Future of High-Sensitivity Charm Experiments}, {\it op cit.}, p.\ 221.

\bibitem{Karchin}
C. J. Kennedy, R. F. Harr, and P. E. Karchin,
``Simulation Study of a Transverse Energy Trigger for a
Fixed Target Beauty and Charm Experiment at Fermilab,"
Sept. 9, 1993 (unpublished).

\bibitem {Processors}
C. Lee {\it et al.}, IEEE Trans. Nucl. Sci. {\bf 38} 461, (1989);
M. Adamovich {\it et al.}, {\it ibid.} {\bf 37} No. 2, 236
(1990); B. C. Knapp,
Nucl. Instr. \& Meth. {\bf A289}, 561 (1990).

\bibitem{Selove}
W. Selove, in {\sl Proceedings of the Workshop on $B$ Physics at Hadron
Accelerators}, P. McBride and C. S. Mishra, {\it eds.}, Fermilab-CONF-93/267
(1993), p. 617.

\bibitem{optrig} G. Charpak, L. M. Lederman, and Y. Giomataris,
Nucl. Instr.
\& Meth. {\bf A306}, 439 (1991);
D. M. Kaplan {\it et al.}, {\it ibid.} {\bf A330}, 33 (1993);
G. Charpak {\it et al.}, {\it ibid.} {\bf A332}, 91 (1993).

\bibitem{mul-jump}
A. M. Halling and S. Kwan, Nucl. Instr. \& Meth. {\bf A333}, 324 (1993).

\bibitem{P865}
L. D. Isenhower {\it et al.},
``P865: Revised Letter of Intent
for a High-Sensitivity Study of Charm and Beauty Decays," April 2, 1993.

\end{thebibliography}
\end{document}